\documentclass[12pt,preprint]{aastex}

\begin{document}
\newcommand{\pul}{PSR~B1706-44}
\newcommand{\xmm}{{\it XMM-Newton}}
\newcommand{\chandra}{{\it Chandra}}

\shorttitle{PSR~B1706-44}
\shortauthors{McGowan et al.}

\title{{\it XMM-Newton} Observations of PSR~B1706-44}
\author{K.E. McGowan\altaffilmark{1}, S. Zane\altaffilmark{2},
M. Cropper\altaffilmark{2}, J.A. Kennea\altaffilmark{3}, 
F.A. C\'{o}rdova\altaffilmark{3}, C. Ho\altaffilmark{1}, \\ 
T. Sasseen\altaffilmark{4}, W.T. Vestrand\altaffilmark{1}}
\altaffiltext{1}{Los Alamos National Laboratory, MS D436, Los Alamos, NM 87545}
\altaffiltext{2}{Mullard Space Science Laboratory, University College of
London, UK}
\altaffiltext{3}{University of California, Riverside, CA 92521}
\altaffiltext{4}{University of California, Santa Barbara, CA 93106}
\email{mcgowan@lanl.gov}

\begin{abstract} 

We report on the \xmm\ observations of the young, 102~ms pulsar
\pul. We have found that both a blackbody plus power-law and a
magnetized atmospheric model plus power-law provide an excellent fit
to the EPIC spectra. The two scenarios are therefore indistinguishable
on a statistical basis, although we are inclined to prefer the latter
on physical grounds. In this case, assuming a source distance of $\sim
2.3$~kpc, the size of the region responsible for the thermal emission
is $R \approx 13$~km, compatible with the surface of a neutron star.
A comparison of the surface temperature of \pul\ obtained from this
fit with cooling curves favor a medium mass neutron star with $M \sim
1.45 M_\sun $ or $ M \sim 1.59 M_\sun$, depending on two different
models of proton superfluidity in the interior.  The large collecting
area of \xmm\ allows us to resolve a substructure in the broad soft
X-ray modulation detected by \chandra, revealing the presence of two
separate peaks with pulsed fractions of $7\pm 4\%$ and $15\pm 3\%$,
respectively.  

\end{abstract}

\section{INTRODUCTION}
\renewcommand{\thefootnote}{\fnsymbol{footnote}}
\setcounter{footnote}{0}

\pul\ is a young ($\tau = 1.7 \times 10^4$~yr), energetic 102~ms
pulsar originally discovered by \citet{john92}.  It is one of
several sources with spin down ages 10$^{4}$--10$^{5}$~yrs which are
referred to as Vela-like pulsars, due to their similar
emission properties (\citealt{bp02}). The source is known to display 
glitches (\citealt{jm95}) and is plausibly associated with the
supernova remnant G343.1-2.3 (\citealt{mc93}; \citealt{dod02};
\citealt{bock02}). 

According to the new \citet{cl02} model, the pulsar is $\sim 2.3$~kpc
away, which agrees with the kinematic distance in the range
2.4--3.2~kpc inferred from H~I absorption (\citealt{koribalski95}).
VLA images (\citealt{fr94}, \citealt{giac01}) indicated that the
pulsar is located inside a synchrotron plerionic nebula about 
$3\arcsec .5 \times 2\arcsec .5$ in size. Evidence for a more extended
X-ray compact nebula (with radius $\sim20\arcsec -30\arcsec$) was also
found in {\it ROSAT}-HRI and \chandra\ images (\citealt{fin98}, 
\citealt{dod02}).

\pul\ is one of only eight radio pulsars which are known to emit in
the GeV range (\citealt{thom92}), and one of only three detected in
the TeV range, although this detection is marginal (\citealt{kifu95}, 
\citealt{ch98}). \citet{thom92} found pulsations at the radio period
in the {\it EGRET}-CGRO data.  The light curve above 400~MeV is
complex, with evidence of two, possibly three peaks, none of which are
in phase with the radio peak.  We note that the light curve of Vela in
the gamma-ray range shows two peaks (\citealt{kan94}), and at least
three in the X-rays (\citealt{stri99,pav00}).  Particular
characteristics of the Vela-like pulsars, with respect to the
Crab-like pulsars, are that the pulse profiles at different energies
are shifted in phase with respect to each other.

An unpulsed X-ray source at the radio-pulsar position had been
detected with {\it ROSAT}-PSPC (\citealt{b95}), ASCA (\citealt{fin98})
and BeppoSax (\citealt{mineo02}).  More recently, deeper \chandra\ 
observations have been presented by Gotthelf, Halpern \& Dodson
(2002).  These authors discovered a broad sinusoidal X-ray pulsation
at the radio-period, with a pulsed fraction of $23\% \pm 6\%$. The
phasing of the radio pulse was consistent with that of the center of
the broad X-ray peak. The high spectral and spatial resolution of
\chandra\ allowed a multi-component fit of the X-ray spectrum
revealing the presence of a thermal component: the X-ray spectrum was
found to be well-fit with a two component model, with a blackbody
temperature of $T=(1.66_{-0.15}^{+0.17}) \times 10^{6}$~K and a power
law index of $\Gamma = 2.0\pm 0.5$. The blackbody radius
$R=3.6\pm0.9$~km determined from the model parameters, suggests that
the emission is from a hot spot, or that a fit with an atmospheric
model is required (\citealt{gott02}). Hydrogen atmospheres are in
general harder than a blackbody since free-free absorption dominates
in the X-ray band. Therefore, the temperature fitted by a hydrogen
atmosphere is lower than that resulting from a blackbody fit (hence
the former fit yields a larger radius).  

In this paper we report on \xmm\ observations of \pul.  Results
from the spectral and timing analyses are presented in the 
following sections. 

\section{OBSERVATIONS}

\pul\ was observed with \xmm\ for $\sim40$~ks on 2002 March 12,
and $\sim46$~ks on 2002 March 13, as a part of the Guaranteed Time
program.  The following analysis uses data from the three European
Photon Imaging Camera (EPIC) instruments: two EPIC MOS detectors
(\citealt{tur01}) and the EPIC PN detector (\citealt{stru01}).  The thin 
optical blocking filter was used on the PN.  To minimize pile-up the PN was
operated in {\it small window} mode, which gives a temporal resolution
of 6~ms.  The MOS1 was operated in {\it full window} (imaging)
mode with a time resolution of 1.4~s.  In order to obtain better
temporal resolution the MOS2 was operated in {\it timing mode}; in
this mode data from the central CCD are collapsed into a
one-dimensional row to achieve a 1.5~ms time resolution.  The medium
filter was used for both MOS observations.  Observations of \pul\ were
taken with the Resolution Grating Spectrometer (RGS), however we did
not detect enough photons for a meaningful analysis.  Due to an
optically bright source in the field the blocking filter was used on
the Optical Monitor (OM).

We reduced the EPIC data with the \xmm\ Science Analysis System (SAS
v5.4.1).  To maximize the signal-to-noise we filtered the data to
include only single, double, triple and quadruple photon events for
MOS, and only single and double photon events for PN.  We also
filtered the events files to exclude times of high background.

\begin{figure}[h] 
\vbox to6in{\rule{0pt}{6in}}
\includegraphics{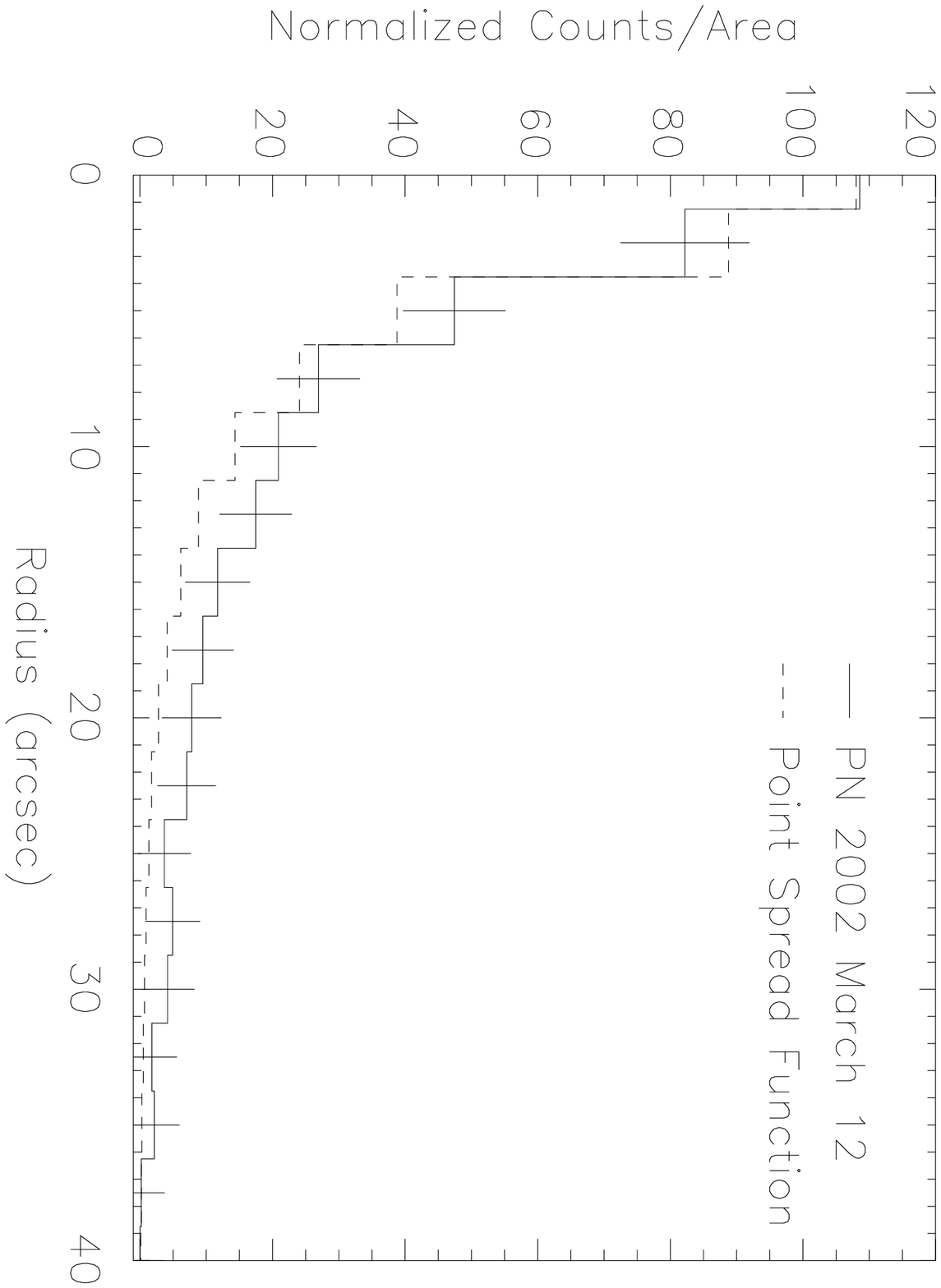} 
\includegraphics{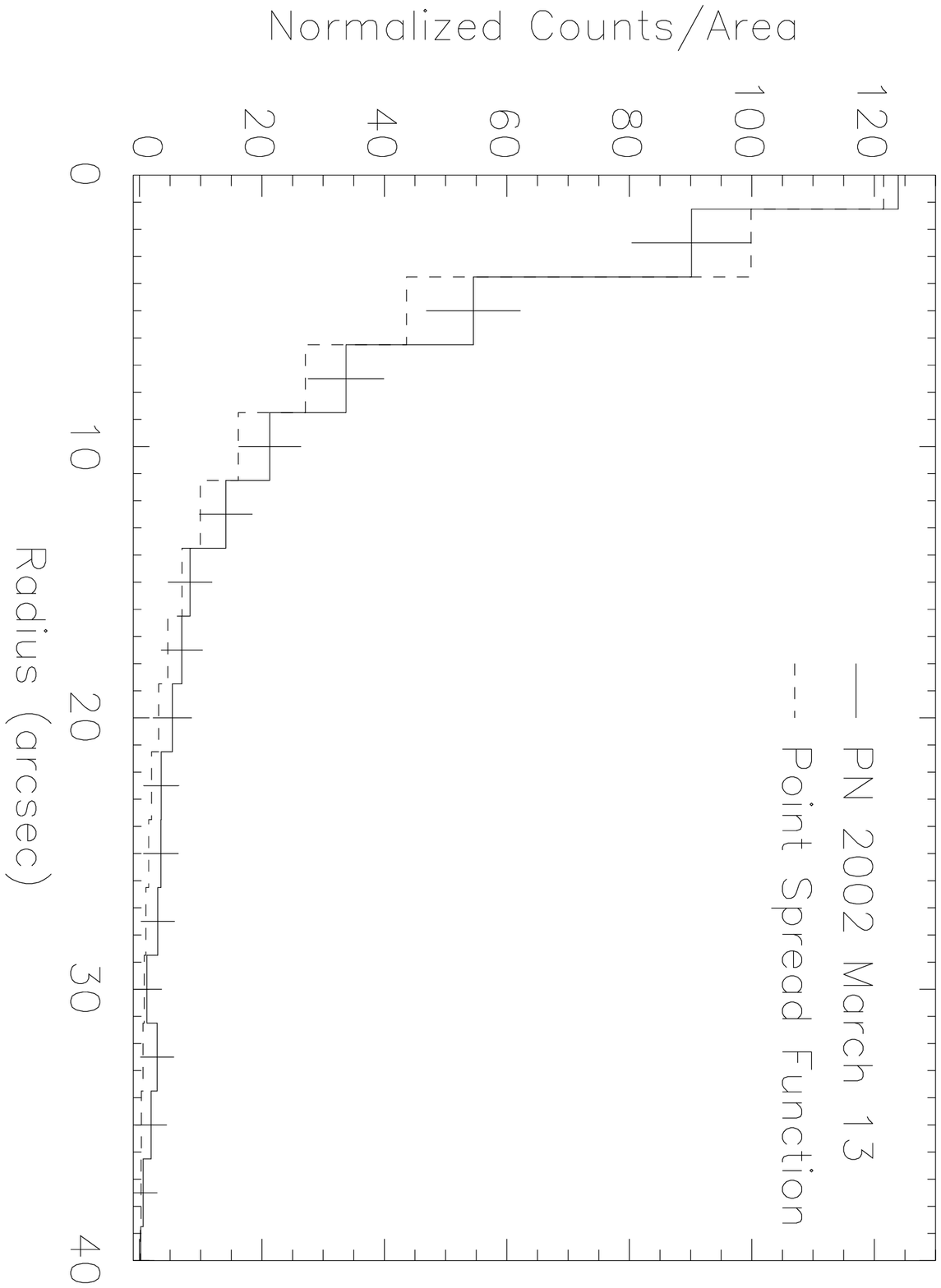} 
\caption{X-ray emission from \pul\ as a function of radius 
({\it solid line}), compared to emission from a simulated point source 
({\it dashed line}; see text).  PN observations from 2002 March 12 
({\it top}) and 13 ({\it bottom}).}\label{fig:radial} \end{figure}

\pul\ is known to be surrounded by a pulsar wind nebula
(\citealt{gott02,dod02}).  In an ideal situation analysis of the
source should be carried out on data which are not contaminated by the
diffuse emission. However, the nebula is very compact, with the
diffuse emission visible between $1\arcsec$ and $10\arcsec-20\arcsec$
(see \citealt{gott02}). Therefore the spatial resolution of \xmm\ is
not high enough to be able to separate the pulsar from its putative
synchrotron nebula. In order to quantify this effect, we have compared
\pul's emission with that for a point source. We simulated the
expected PN data for a point source with similar column depth and
spectral characteristics as \pul\ using {\it
SciSim}\footnote{http://xmm.vilspa.esa.es/}.  The radial profiles for
\pul\ from 2002 March 12 and 13, and the simulated source, are shown
in Fig.~\ref{fig:radial}.  The background has been subtracted from
both the \pul\ and point source data.  We have normalized the total
point source counts to the total counts from \pul\ by performing a
$\chi^{2}$ minimization to determine the appropriate scaling factor
required to fit only the core of the pulsar ($<5\arcsec$).  

Our results show marginal evidence for an excess of diffuse emission
between $5\arcsec$ and $20\arcsec$.  An extraction radius of
$<5\arcsec$ does not give a sufficient number of photons for analysis
purposes, hence, we have chosen an extraction radius of $20\arcsec$.
We note that this radius will therefore include emission from the pulsar
wind nebula. As only $\sim75$~\% of the energy from the source is
encircled in this radius the measured fluxes have been corrected in
the following analyses.

\section{SPECTRAL ANALYSIS}

Spectra for \pul\ were extracted from both the PN and MOS1 data, and
regrouped by requiring at least 30 counts per spectral bin.  We
subtracted a background which was extracted from an annulus around the
source fiducial region.  The corresponding photon redistribution
matrix (RMF) and ancillary region file (ARF) were created.  

\begin{figure}[h]
\epsscale{0.8}
\plotone{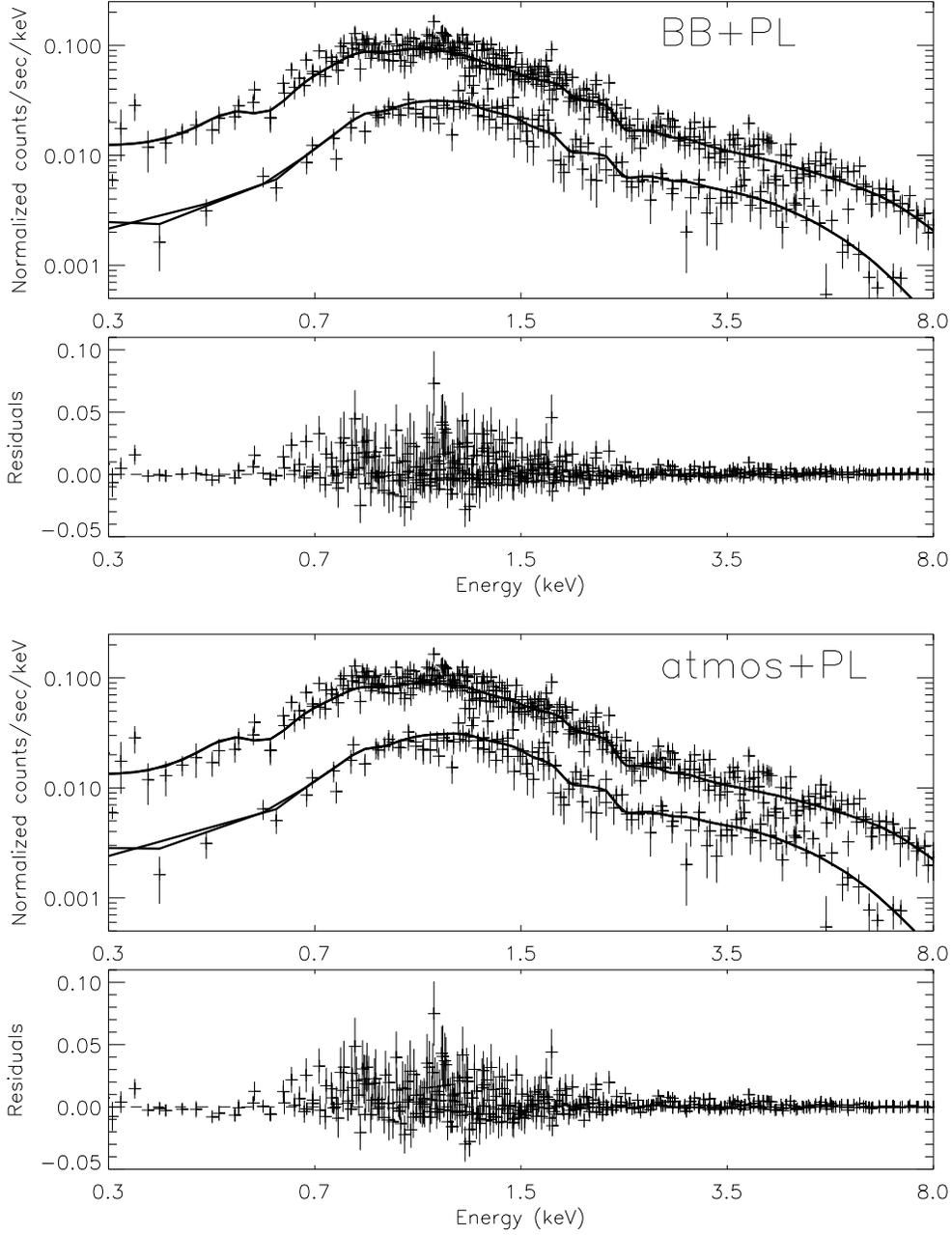}
\caption{PN and MOS1 spectra of \pul.  First panel, data ({\it crosses})
and best-fit blackbody plus power-law model with free $N_{H}$ 
({\it thick line}) for the parameters given in Table~\ref{spec}.
Second panel, the difference between the data and the blackbody plus 
power-law model.  Third panel, data ({\it crosses}) and best-fit
magnetic atmosphere plus power-law model ({\it thick line}) for the 
parameters given in Table~\ref{spec}, where the radius of the neutron
star was fixed at 12~km.  Second panel, the difference between the
data and the magnetic atmosphere plus power-law model.}\label{fig:spec}
\end{figure}

We simultaneously fit the four spectra, two from the PN and two from
MOS1, with thermal and power-law models, modified by photoelectric
absorption.  Results are shown in Fig.~\ref{fig:spec} and 
the best-fit parameters are summarized in Table~\ref{spec}.  The
column density was first fixed at the value given in \citet{gott02}, 
$N_{H}=5.5 \times 10^{21}$~cm$^{-2}$, for the single blackbody, single
power-law and blackbody plus power-law fits.  As this value was not
very well constrained in \citet{gott02}, we also allowed $N_{H}$ to
vary for the above model fits.  We find that the data are poorly fit
with a single blackbody for free and fixed $N_{H}$, and with a single
power-law with fixed $N_{H}$.  The single power-law model with free
$N_{H}$ fits the data reasonably well, however the two-component
blackbody plus power-law model with either free or fixed $N_{H}$ gives
a better fit.  We find the two-component fit where the column density
is allowed to vary results in $N_{H}=(4.5_{-0.4}^{+0.7}) \times
10^{21}$~cm$^{-2}$, with a temperature
$T^{\infty}=(2.01_{-0.20}^{+0.18}) \times 10^{6}$~K, power-law index
$\Gamma = 1.49_{-0.08}^{+0.09}$, and $\chi^{2}_{\nu}=0.84$ for 658
d.o.f.  The X-ray luminosity is $1.0 \times 10^{33}$ ergs s$^{-1}$ in
the $0.2-10$~keV band, and the blackbody contribution is $3.8\times
10^{32}$ ergs s$^{-1}$.  The distance to \pul\ is uncertain, with
values in the range $1.8-3.2$~kpc (\citealt{taylor93,koribalski95}).
By adopting the value determined from the \citet{cl02} free-electron
model of the Galaxy, $D=2.3\pm 0.3$~kpc, we obtain an emitting radius
$R^{\infty}=1.81_{-0.29}^{+0.43}$~km, too small to be compatible with a 
neutron star equation of state.  The results for the column density,
temperature and power-law index are in good agreement with those
obtained by \citet{gott02} based on \chandra\ data.  While the radius
we determine is smaller than that from the fits to the \chandra\ data,
\citet{gott02} also found that the radius of the emitting region
inferred from the blackbody fit of the thermal component indicates a
hot region on a cooler surface.

In order to try a more physical description of the thermal component, 
we fitted the data with a neutron star hydrogen atmosphere plus power-law 
model.  The magnetic field of \pul, as inferred from the spin-down rate, is 
$3\times 10^{12}$~G, high enough to have a substantial effect on the 
opacities (\citealt{pav92,zav96,zane00}). We used a grid of pure-H, 
atmospheric cooling models computed for $B=10^{12}$~G and different 
effective temperatures, provided by V.~Zavlin (V.E. Zavlin 2003,
private communication). 
 
We first fixed the neutron star mass and radius at $M_{NS}=1.4
M_\sun$, $R_{NS}=10$~km; all other parameters were allowed to vary.
The resulting best-fit parameters are $\Gamma=1.45_{-0.01}^{+0.14}$,
$N_{H}=(5.2\pm 0.1)\times 10^{21}$~cm$^{-2}$ and
$T_{eff}=(1.03_{-0.31}^{+0.07})\times 10^{6}$~K, where $T_{eff}$ is
the effective temperature evaluated at the star surface (see
Table~\ref{spec}).  This gives a temperature at infinity $T^\infty =
(0.79_{-0.31}^{+0.07})\times 10^{6}$~K.  The source distance resulting
from this fit is $D=1.7\pm 0.3$~kpc, lower than that computed by
\cite{cl02}.  The reliability of the pulsar distance derived using the
dispersion measure method is uncertain, however, the distance issue
can only be firmly addressed by parallax measurements, which are not
available for \pul.  We therefore fit the data assuming a distance to
the source of $2.3\pm0.3$~kpc.  By fixing $R_{NS}=12$~km and repeating
the atmospheric plus power-law model fit we find
$\Gamma=1.43_{-0.05}^{+0.20}$, $N_{H}=(5.1_{-0.1}^{+0.2})\times
10^{21}$~cm$^{-2}$ and $T_{eff}=(1.01_{-0.34}^{+0.01})\times 10^{6}$~K
[$T^\infty=(0.82_{-0.34}^{+0.01})\times 10^{6}$~K].  The distance to
the pulsar from this fit is $D= 2.1\pm 0.2$~kpc, in agreement with the
value obtained from the dispersion measure.  

Due to the goodness of fit for both the blackbody plus power-law, and the 
magnetic atmosphere plus power-law models we are unable to distinguish 
between these two scenarios on only a statistical basis.  Taken at
face value, these results indicate that the emission is either from
the whole neutron star surface with radius $\sim 12$~km, or that the
thermal X-rays originate from a smaller region, i.e. a hot spot. 

\section{TIMING ANALYSIS}

For the timing analysis the event files were filtered on energy, to
include only photons in the range $0.2-10$~keV. The filtered
event files were then barycentrically corrected.  We extracted data
for the PN from a circular region centered on the source of radius 
$20\arcsec$.  Selection of the photons from the MOS2 event files was
achieved by extracting 90\% of the flux within a rectangular region
centered on the source.  The fluxes have been corrected for this in
the following temporal analysis.  To increase the signal-to-noise we
have combined the PN data from 2002 March 12 and 13.  We also combined
the MOS2 datasets.     

We determine a predicted pulse period at the epoch of our \xmm\
observations for \pul, assuming a linear spin-down rate and using the 
radio measurements (\citealt{wang00,gott02}).  We find
$P=102.477638$~ms ($f=9.7582265$~Hz) at the start of our first
observation (MJD 52345.0), and $P=102.477654$~ms ($f=9.7582249$~Hz) at
the end of our second observation (MJD 52347.0).  As the frequency of
the predicted period varies over the duration of our observations, and 
glitches and/or deviations from a linear spin-down may alter the
period, we searched for a pulsed signal in the PN and MOS2 data over a
wider frequency range centered on $f=9.75823$~Hz.

We have employed two methods in our search for pulsed emission from
\pul.  In both methods we have included the frequency derivative from
\cite{gott02} in our calculations to determine the best-fit frequency.
In the first method we implement the $Z^{2}_{n}$ test
(\citealt{bucc83}), with the number of harmonics $n$ being varied from
1 to 5.  In the second method we calculate the Rayleigh Statistic (see
\citealt{deJager91,mardia72}) and then calculate the Maximum
Likelihood Periodogram (MLP) using the $\Delta C$-statistic
(\citealt{cash79}) to determine significant periodicities in the
datasets (see \citealt{zane02}). 

\begin{figure}[h]
\vbox to5in{\rule{0pt}{5in}}
\includegraphics{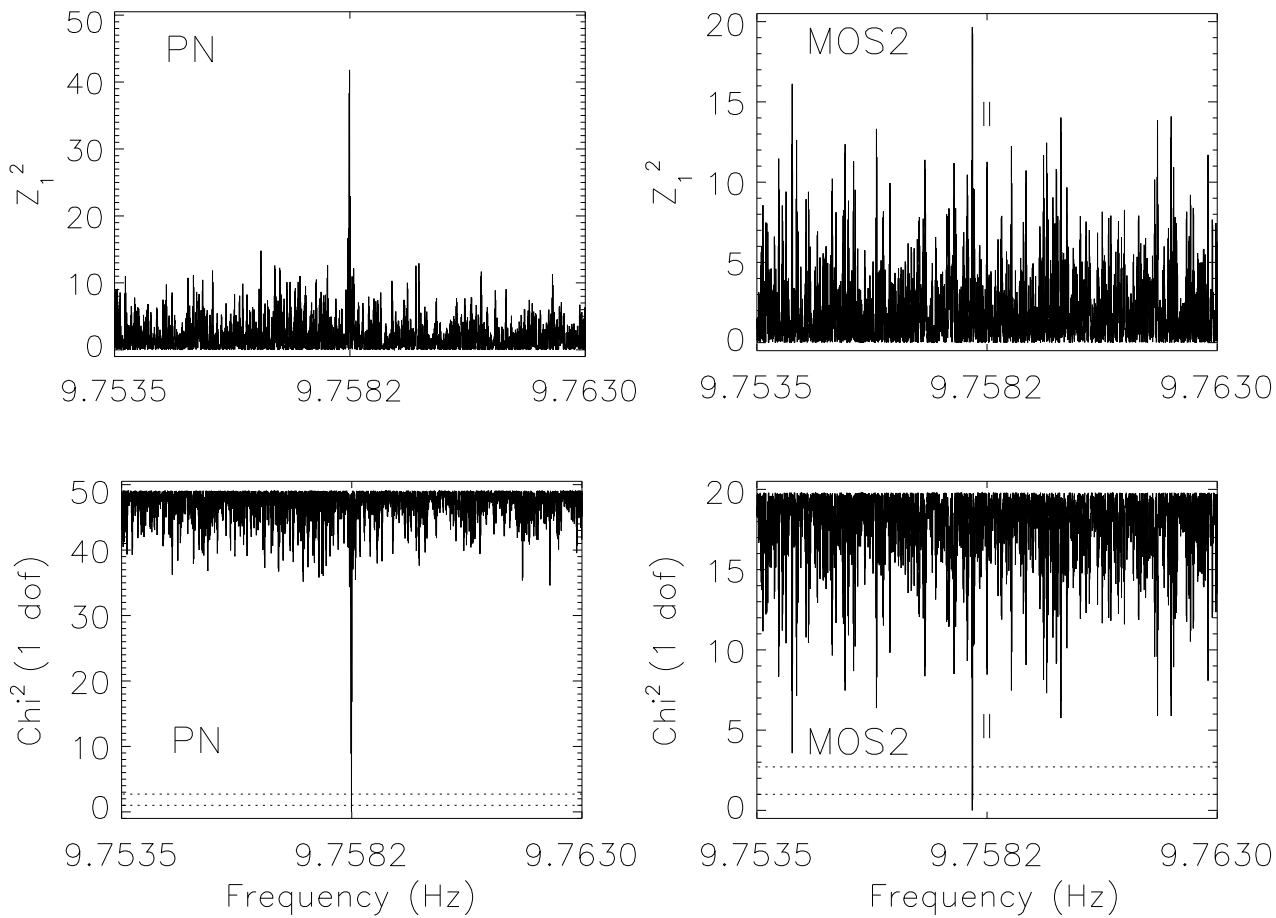}
\caption{$Z^{2}_{1}$ ({\it top}) and maximum likelihood periodograms
({\it bottom}) for the combined PN ({\it left}) and MOS2 data ({\it
right}) of \pul.  The highest peak in the $Z^{2}_{1}$ periodogram for
the PN data occurs at $9.7582258$~Hz, with the corresponding MLP peak
at $9.7582263$~Hz.  The peak nearest to the predicted frequency in the
MOS2 $Z^{2}_{1}$ periodogram and MLP occurs at $f=9.7582258$~Hz and is
marked by two {\it solid lines}.  The 68\% and 90\% confidence levels
for the periods in the MLP are at $\chi^{2}=1.0$ and $2.71$ ({\it dotted lines}).}\label{fig:psearch} 
\end{figure}

The most significant $Z^{2}_{n}$-statistic for the combined PN data
occurs for $n=1$.  With the number of harmonics equal to one, the
$Z^{2}_{n}$-statistic corresponds to the well known Rayleigh
statistic.  We find a peak at
$f=9.7582258_{-0.0000009}^{+0.0000007}$~Hz (see
Fig.~\ref{fig:psearch}, {\it top left panel}).  The quoted 90\%
uncertainty range is determined from the error in the position of the 
peak.  Within errors, the peak in the $Z^{2}_{1}$ periodogram lies in
the range of predicted frequencies.  The $Z^{2}_{1}$-statistic for
this peak is 41.82, which has a probability of chance occurrence of 
$8.3\times 10^{-10}$.  The corresponding peak in the MLP periodogram
occurs at $f=9.7582263_{-0.0000005}^{+0.0000001}$~Hz
(Fig.~\ref{fig:psearch}, {\it bottom left panel}).  Within the 90\%
confidence limit, this value is consistent with the predicted radio
frequency, and the frequency found from the $Z^{2}_{1}$ periodogram.

We generated $Z^{2}_{n}$ and MLP periodograms for the MOS2 data as for
the PN data.  There is a peak at $f=9.7582258$~Hz in the $Z^{2}_{1}$
and MLP periodograms, however it is not the most significant peak.
There are several peaks present in both periodograms, the significance
of which are low compared to the noise level.  The individual PN
datasets from 2002 March 12 and 13 have $S/N=5.0$ and $S/N=6.5$,
respectively.  The MOS2 data from the same dates both have $S/N=1.6$.
The low signal-to-noise ratio of the MOS2 data is the cause for the
lack of a significant peak at the predicted frequency.   

\section{FOLDED LIGHT CURVE}

As our \xmm\ observations were taken more than a year outside the
valid range of the radio ephemeris given in \citet{gott02} we are
unable to determine a phase relationship between the X-ray and
radio pulse profiles.  Hence, we have used an arbitrary ephemeris
$T_{0}=$~MJD~$52346.2$.  We folded the PN and MOS2 data on the frequency
found from the $Z^{2}_{1}$ periodogram, $f=9.7582258$~Hz.  The effects
of the drifting period are taken into account by including the
frequency derivative in the calculations.  

\begin{figure}[h]
\epsscale{0.9}
\plotone{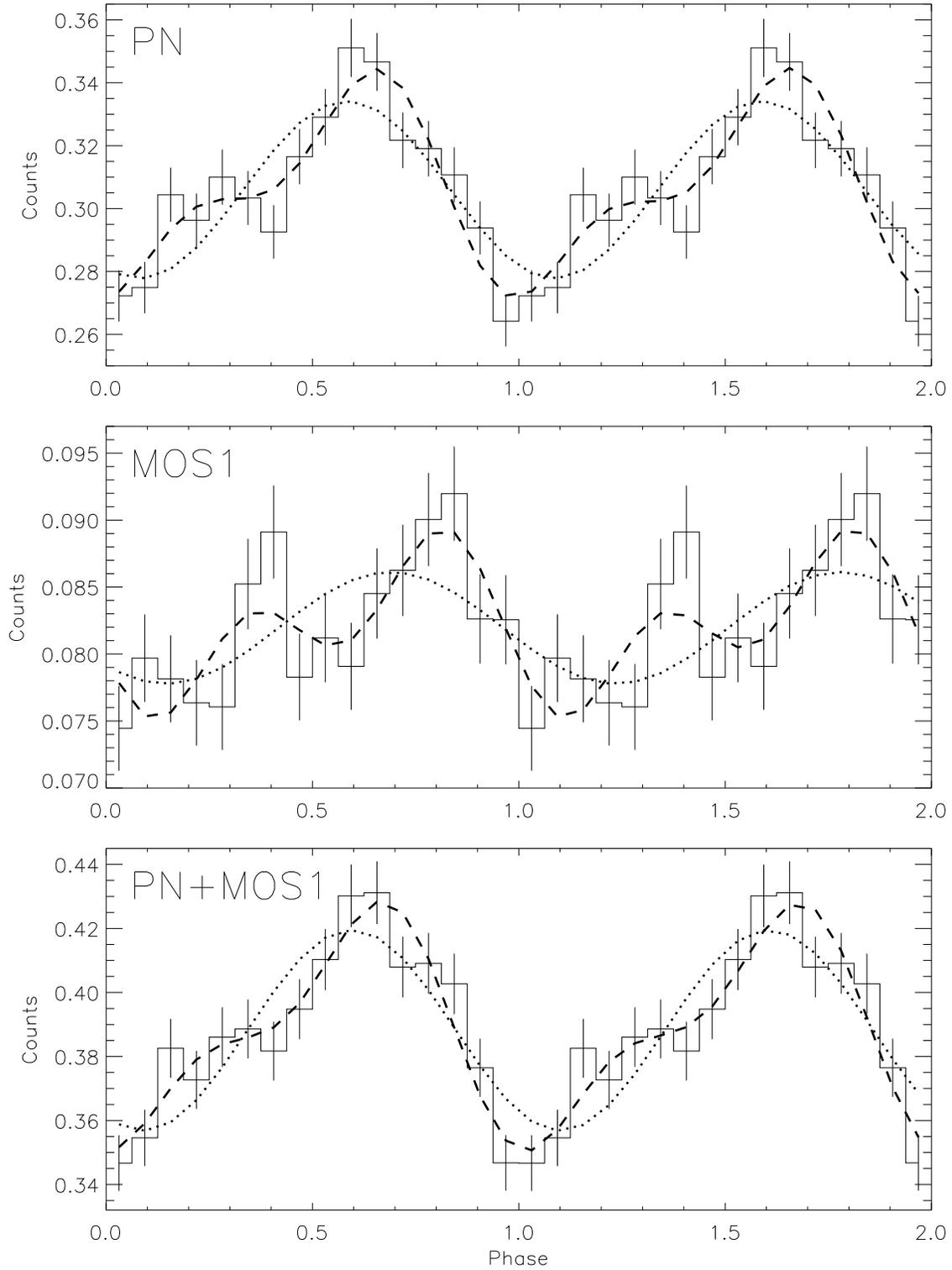}
\caption{Pulse profiles for \pul.  Data are folded using
$f=9.7582258$~Hz.  {\it Top panel}, combined PN data; {\it middle
panel}, combined MOS2 data; {\it bottom panel}, combined PN and MOS2
data.  Best-fit models are shown, one sinusoid ({\it dotted line}),
two sinusoids ({\it dashed line}).}\label{fig:prof}
\end{figure}

In Fig.~\ref{fig:prof} we show the combined PN, MOS2 and PN plus MOS2
pulse profiles.  We modeled the folded light curves with one and two 
sinusoids; the resulting $\chi^{2}_{\nu}$ values are given in 
Table~\ref{sinfit}.  As it can be seen, the fits to the data are
improved with the double sinusoidal model.  To determine the
statistical significance of the second sinusoid we employed an F-test.
We find F-test probabilities of $9\times 10^{-5}$, $3\times 10^{-3}$
and $4\times 10^{-5}$, for the PN, MOS2 and PN plus MOS2 fits
respectively.  These values indicate that the PN and PN plus MOS2 data
are best-fit with the two sinusoid model, while no firm conclusion can
be derived from the low S/N MOS2 data.  The pulsed fractions of the
two peaks are $7\pm 4\%$ and $15\pm 3\%$.  

\begin{figure}[h]
\epsscale{0.8}
\plotone{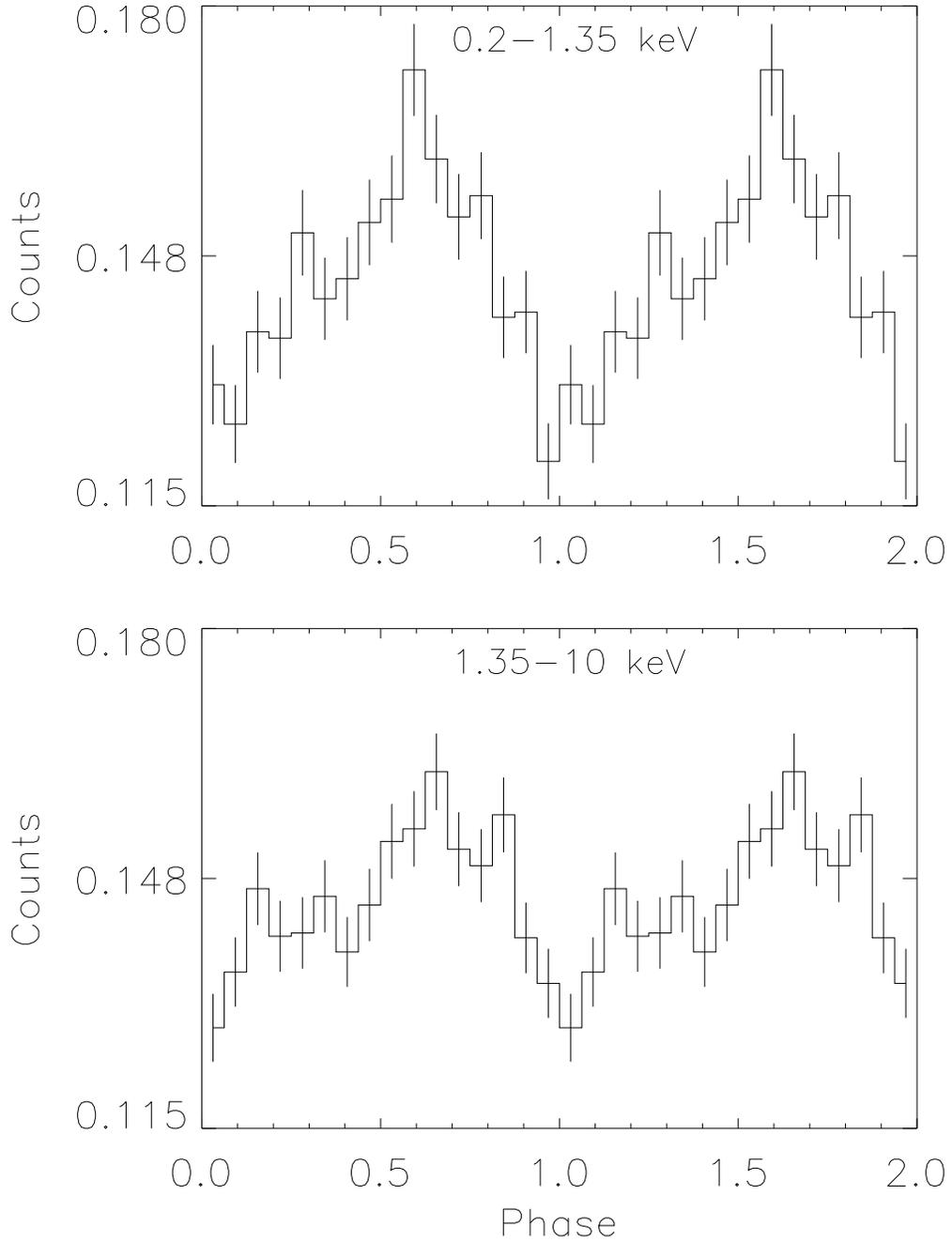}
\caption{Pulse profiles for combined PN and MOS2 data for \pul\ in
different energy ranges.  Data are folded using $f=9.7582258$~Hz.  
{\it First panel}, $0.2-1.35$~keV; {\it second panel}, $1.35-10.0$~keV.}\label{fig:diffe}
\end{figure}

In an attempt to investigate if the two peaks in the folded light 
curve are thermal or non-thermal in origin, we determined from the 
two-component model fits to the spectra the energy at which the
power-law starts to dominate over the thermal emission.  This occurs
at $\sim1.23$~keV, and $\sim1.34$~keV in the blackbody plus power-law
and magnetic atmosphere plus power-law models, respectively.  By
filtering the PN and MOS2 event files on energy we produced light
curves in the following energy ranges $0.2-1.35$ and $1.35-10.0$~keV,
which we barycentrically corrected.  We folded the data on
$f=9.7582258$~Hz (Fig.~\ref{fig:diffe}), again including the frequency
derivative in the calculations.  As we can see, the same features are
present in both bands; the thermal and non-thermal emission are both 
pulsating and phase aligned.  However, at higher energies the
modulation is lower, indicating that the power law component is
contaminated by the non-pulsating nebular emission.

\section{DISCUSSION}
\label{sect:disc}
\subsection{Spectral Results}
\label{sect:discsp}

Of the $>1000$ radio pulsars detected so far, only $<5\%$
have also been found in the X-rays. These X-ray emitting pulsars
represent a wide range of ages ($10^3-7\times 10^9$~yrs), magnetic
field strengths ($10^8-10^{13}$~G), periods ($1.6-530$~ms) and
spectral properties. In particular, only a sub-set of them are
suitable for observing the thermal radiation from the neutron star 
surface, and therefore constrain the atmospheric chemical
composition and the pulsar cooling history.  Thermal emission from the 
neutron star surface is not detectable in pulsars older than
$10^{6}$~yr: standard cooling scenarios predict a sharp reduction in
the surface temperature when surface photon emission overtakes the
neutrino luminosity losses (\citealt{nt87}). In pulsars younger than
$10^{4}$~yr strong non-thermal emission from the magnetosphere or the
synchrotron nebula swamps the weaker thermal radiation and dominates
the X-ray spectrum. Only in middle-aged pulsars (ages $10^4-10^6$~yr),
is the non-thermal component much fainter; hence for such objects the
thermal radiation from the neutron star surface ($T\sim 0.02-0.1$~keV)
can dominate at soft X-ray/UV energies.  

To date, thermal emission has been detected in only very few
radio pulsars, e.g.\ PSR~B0656+14 (\citealt{po96}), PSR~B1055-52
(\citealt{pa02}), PSR~J0437-4715 (\citealt{zavlin02}), PSR~J0538+2817
(\citealt{mcgowan03}), Geminga (\citealt{hw97}), Vela
(\citealt{pa01}), and \pul\ (\citealt{gott02}).  The thermal emission
detected above $\sim 0.5$~keV in the spectrum of the first four
objects has been more plausibly interpreted as originating from a
hot-polar cap. This agrees with the fact that these older sources
should have a surface flux peaked in the UV and it is confirmed by a
detection, in the brightest PSR~B0656+14, of a further thermal
component below $0.7$~keV (\citealt{pa02}). The younger Vela pulsar 
($\tau = P/2 \dot P  \sim 10^4$~yr) is the only radio active source 
for which the thermal component observed in the soft X-rays is well
explained by a magnetized cooling atmosphere (\citealt{pa01}). When
this model is assumed instead of a blackbody, the inferred radius
increases from $R^{\infty}/d_{294} \approx 2.5 $~km to
$R^{\infty}/d_{294} \approx 15^{+6.5}_{-5.1}$~km (where $d_{294}$ is
the distance in units of 294~pc) and is therefore in agreement with a
standard neutron star equation of state.  The only other neutron stars
whose thermal component is better described by an atmospheric model,
and for which this interpretation resolves all the inconsistencies
which follow from the blackbody interpretation, are the radio-silent
neutron stars 1E~1207-52 (\citealt{za98}) and RX~J0822-4300
(\citealt{za99}).   

Here we present a further example. We have found that both a blackbody
plus power-law and a magnetized atmospheric model plus power-law
provide an excellent fit to the data, and are indistinguishable on a
statistical basis. However, the latter has to be preferred on physical
grounds, as we argue below, and, at a source distance of $2.3$~kpc,
gives an emitting area of $R = 13.41^{+1.75}_{-4.84}$~km, compatible
with the size of a neutron star (\citealt{lt01}).

Our knowledge of neutron star interiors is still uncertain and accurate
measurements of the neutron star surface temperature are particularly
important to constrain the cooling models and provide information on the
physics of the neutron star. Roughly speaking, theoretical models predict
a two-fold behavior of the cooling curves. In low-mass neutron stars
neutrino emission is mainly due to a modified Urca process and
nucleon-nucleon bremsstrahlung. These are relatively weak mechanisms and
produce {\it slow cooling}. In stars of higher mass the neutrino emission
is enhanced by a direct Urca process (or other mechanisms in exotic
matter), therefore these stars cool down much faster ({\it fast cooling}
regime). To date (see \citealt{ya02} for a recent a discussion) it has
been realized that simple models which do not account for proton and
neutron superfluidity fail in explaining the surface temperatures observed
in many sources, unless objects such as e.g. Vela, Geminga, RX~J1856-3754
do have exactly the critical mass that bounds the transition between
the very different {\it slow cooling} and {\it fast cooling} regimes. This
unlikely assumption can be avoided by including the effects of nucleon
superfluidity. Models with proton superfluidity included predict an
intermediate region between fast cooling and slow cooling curves, which is
expected to be populated by medium mass neutron stars (roughly with $M$
between 1.4 and 1.65 $M_\sun$). Although the full picture only holds if,
at the same time, neutron superfluidity is assumed to be rather weak, it
is still interesting that many neutron stars (as 1E~1207-52, Vela,
RX~J1856-3754, PSR~0656+14)  have a surface temperature which falls in
such a transition region. Since the cooling curves in the transition
region show a significant spread with the neutron star mass, if this
scenario is correct we can select those curves which explain the
observations and therefore attribute certain masses to the sources
(``weighing'' neutron stars, \citealt{ka01}). As we can see from the
first two panels of Figure~2 in \cite{ya02}, assuming an age of $\log \tau
= 4.23$, the surface temperature of \pul \ derived from the blackbody fit
is even higher than the upper cooling curves i.e. those corresponding to
the slow cooling regime. However, the surface temperature $\log T^\infty =
5.9$ obtained by fitting with the magnetized model and $R=12$~km falls
well within the above mentioned transition region of medium mass neutron
stars. The mass of \pul \ should then be $\sim 1.45 M_\sun $ or $ \sim
1.59 M_\sun$, depending on the kind of proton superfluidity assumed in the
model (1p and 2p respectively).  We note that interpreting the
temperatures obtained from the spectral fits in the context of
theoretical cooling curves relies on the true age of the pulsar being
the same as the characteristic spin-down age, which may not be valid.

\subsection{Timing Results}
\label{sect:disctim}

The first detection of pulsations in the soft X-rays of \pul\ by 
\chandra\ was a broad modulation in the light curve, approximately in
phase with the radio peak (\citealt{john92,gott02}).  Its is known
that the radio and gamma-ray pulses are not aligned, with the
gamma-ray emission occurring $\sim 0.37$ cycles after the radio
emission (\citealt{th96}).  The \xmm\ data allows us to resolve the 
substructure of the soft X-ray light curve, revealing the presence of
two separate peaks.  However, due to the lack of a contemporaneous
radio observation of this glitching pulsar, we are unable to determine
an absolute phase relationship between the X-ray, radio and gamma-ray
pulse profiles.  

To investigate the origin of the pulsed emission in \pul\ it is
instructive to consider the timing behavior of Vela, since both
sources have similar spectral properties.  The complex multiwavelength
light curves of Vela are the subject of continued interest and some
evidence of correlations between the different energy bands have
recently been established (\citealt{har02}).  Unfortunately, in the
case of Vela, the key to deciphering pulsar emission mechanisms was
the light curve in the $2-30$~keV {\it RXTE} energy band (obtained
with a 92~ks and a $\sim 300$~ks pointing, \citealt{har02}), while a
132~ks {\it RXTE} observation of \pul\ only gave an upper limit on the
pulsed emission in the $9-18$~keV band (\citealt{ray99}).  

The pulse profile of Vela obtained with {\it EGRET} shows two distinct
narrow pulses, at phase $\sim 0.1$ and $\sim 0.55$, neither aligned
with respect to the radio peak.  Considerable emission has been
detected in the phase interval between the peaks (\citealt{kan94}).  
By contrast, in the {\it EGRET} band \pul\ shows a broader modulation
that extends over the full range of phases between $\sim 0.2-0.6$, and
is likely to consist of two broad peaks separated by $\sim 0.2$
(\citealt{th96}).  No gamma-ray emission is detected outside of the
peaks and, as in Vela, the {\it EGRET} pulses are out of phase with
respect to the radio pulse.  The difference in the gamma-ray light
curves of the two sources is likely to be indicative of a different
beaming geometry, with \pul\ having greater alignment between the
rotational and magnetic axes.  This is also confirmed by
radio-polarization studies (\citealt{gu95}).  An alternative, but 
indistinguishable scenario is in terms of different viewing angles, in
which case \pul\ should be viewed more equator-on than Vela. 

The soft X-ray profile of Vela obtained with \chandra\ shows at least
three peaks (\citealt{pav00}), at phases $\approx 0.1, 0.4, 0.8$.  The
presence of several different pulses may be explained in terms of
non-thermal radiation originating from more than one region in the
pulsar magnetosphere (the soft X-ray pulse profile has a total pulsed
fraction of $\sim8$~\%).  However, only the first two peaks have a
plausible association with {\it EGRET} (and {\it RXTE}, see
\citealt{har02}) peaks at similar phases.  Although this association
is more conclusive for the first peak, if it is the case, this
suggests that the emission is probably caused by non-thermal
radiation.  On the other hand, radiation emergent from a stellar
atmosphere is inherently anisotropic regardless of the local
temperature and magnetic field, and, when combined with a non-uniform
temperature profile at the neutron star surface, is capable of
generating considerable pulsar modulation.  Also, pulsations are
expected if the thermal flux originates from hot polar caps.  In this
respect, the fact that the third soft X-ray peak does not have a
counterpart in the gamma-rays led to speculation that it may be
thermal in origin (\citealt{pav00}).  

In these \xmm\ observations of \pul\ we have found evidence 
for two peaks in the soft X-rays.  Whether there are similarities
between these peaks and those in the soft X-ray light curve of Vela,
and the underlying emitting scenario, is unclear.  We find that for
\pul\ the overall X-ray pulsation is much more prominent at softer 
energies, below $\sim 1$~keV, but unfortunately a comparison between 
the light curves in different energy bands does not allow us to
establish unequivocally the thermal or non-thermal character of the two 
peaks (see Fig.~\ref{fig:diffe}).  In fact, due to the modest spatial 
resolution of \xmm, the power-law component detected in the X-ray
spectrum is probably highly contaminated by the non-pulsating nebula,
which explains the overall decrease of the pulsed fraction at higher
energies.  However, we do observe the modulation in both bands and
there is not a significant energy-dependent change in the phasing. 

As mentioned above, it has been only recently that deep
multiwavelength observations of Vela have successfully proven that the
source has nearly phase-aligned pulse profiles and a spectral
continuity from X-rays to gamma-rays (\citealt{har02}).  This is of
great importance, since the connection between the pulse profiles in
different bands was not proven before in Vela-like pulsars, even at a
phenomenological level.  \pul\ is considerably weaker than Vela and it
lies in the Galactic Center region where the diffuse gamma and X-ray
radiation is more intense.  However, our results indicate that the
soft X-ray pulse profile of \pul\ consists of two peaks, which may be
similar to the Vela light curve in the soft X-ray energies.  To
determine if \pul\ is in fact ``Vela-like'' not only spectrally, but
also temporally, future spectrally resolved X-ray observations with
high throughput in the $2-30$~keV band, with contemporaneous radio
observations, are required. 

\section{ACKNOWLEDGMENTS}

We thank the anonymous referee for helpful comments.  We are grateful
to V.E. Zavlin for providing a grid of magnetized cooling atmosphere
models.  The authors thank Gavin Ramsey and Sergey Trudolyubov for
discussions on timing issues.  This work is based on observations
obtained with \xmm, an ESA science mission with instruments and
contributions directly funded by ESA Member States and NASA.  The
authors acknowledge support from the Institute of Geophysics and
Planetary Physics (IGPP) program at LANL and NASA grants S-13776-G and
NAG5-7714.

\clearpage 

\clearpage

\clearpage

\clearpage

\clearpage

\clearpage
\begin{deluxetable}{lcccccc}
\tablewidth{40pc}
\tablecaption{Spectral fits to \pul
\label{spec}}
\tablehead{
\colhead{Model} & \colhead{$N_{H}$}  & \colhead{$\Gamma$} &
\colhead{$R$} & \colhead{$T^\infty$} & \colhead{$D$} &
\colhead{$\chi^{2}_{\nu}$ [dof]} \\
\colhead{} & \colhead{$\times 10^{21}$ cm$^{-2}$}  & \colhead{} &
\colhead{km} & \colhead{$\times 10^{6}$ K} & \colhead{kpc} & \colhead{}}
\startdata
BB & 5.5 (fixed) & \nodata & $0.75_{-0.04}^{+0.06, a}$ &
$3.28_{-0.12}^{+0.08}$ & $2.3\pm0.3$ (fixed) & 4.88 [661] \\
BB & $0.001_{-0.001}^{+0.058}$ & \nodata & $0.10_{-0.02}^{+0.04,a}$ & 
$8.14_{-1.18}^{+1.14}$ & $2.3\pm0.3$ (fixed) & 2.92 [660] \\
PL & 5.5 (fixed) & $2.45_{-0.05}^{+0.05}$ & \nodata & \nodata &
$2.3\pm0.3$ (fixed) & 1.82 [661] \\
PL & $2.9_{-0.2}^{+0.2}$ & $1.83_{-0.05}^{+0.05}$ & \nodata & \nodata
& $2.3\pm0.3$ (fixed) & 1.17 [660] \\
BB+PL & 5.5 (fixed) & $1.57_{-0.06}^{+0.07}$ & $3.23_{-0.20}^{+0.22, a}$
& $1.76_{-0.06}^{+0.06}$ & $2.3\pm0.3$ (fixed) & 0.84 [659] \\
BB+PL & $4.5_{-0.4}^{+0.7}$ & $1.49_{-0.08}^{+0.09}$ & 
$1.81_{-0.29}^{+0.43,a}$ & $2.01_{-0.20}^{+0.18}$ & $2.3\pm0.3$ (fixed)
& 0.84 [658] \\ 
atmos+PL & $5.2_{-0.1}^{+0.1}$ & $1.45_{-0.01}^{+0.14}$ & 10 (fixed) &
$0.79_{-0.31}^{+0.07}$ & $1.7 \pm 0.3$ & 0.84 [658] \\
atmos+PL & $5.1_{-0.1}^{+0.2}$ & $1.43_{-0.05}^{+0.20}$ & 12 (fixed)
& $0.82_{-0.34}^{+0.01}$ & $2.1 \pm 0.2 $ & 0.84 [658] \\
\tablecomments{
The atmospheric cooling models are computed for $B=10^{12}$~G, pure-H 
chemical composition, and have been provided by V.~Zavlin 
(\citealt{zav03}). The local temperature $T_{eff}$ obtained from the
atmospheric fits have been redshifted to infinity according to 
$T^\infty = T_{eff} \sqrt{1-2GM/Rc^2}$, with $R$ given in the $4^{th}$
column and $M=1.4\ M_\sun$. \\
$^a$ Note that this is the value of the radius redshifted at infinity, 
while the entries in the atmospheric fits are the model parameter $R$,
i.e.~the radius measured at the star surface.}
\enddata
\end{deluxetable}

\clearpage
\begin{deluxetable}{lcc}
\tablewidth{17pc}
\tablecaption{$\chi^{2}_{\nu}$ values for sinusoidal fits to the PN,
MOS2, and PN plus MOS2 data
\label{sinfit}}
\tablehead{
\colhead{Data} & \colhead{1 sinusoid}  & \colhead{2 sinusoids}\\
\colhead{} & \colhead{$\chi^{2}_{\nu}$ [dof]} &
\colhead{$\chi^{2}_{\nu}$ [dof]}}
\startdata
PN & 2.6 [28] & 1.3 [25] \\
MOS2 & 1.7 [28] & 1.1 [25] \\
PN+MOS2 & 2.0 [28] & 0.9 [25]\\
\enddata
\end{deluxetable}

\end{document}